\documentclass[aps,preprintnumbers,superscriptaddress,showpacs,twocolumn]{revtex4}
\usepackage{epsfig}
\usepackage{psfrag}
\usepackage{amsfonts}
\usepackage{graphicx,grffile}
\usepackage{dcolumn}
\usepackage{bm}
\def\ie{ \emph{i.e.,} }

\def\et{ \emph{et al}}

\begin{document}

\title{Heavy quark collisional energy loss in  a nonextensive quark-gluon plasma}

\author{Bing-feng Jiang}
\email{jiangbf@mails.ccnu.edu.cn} \affiliation{ College of Intelligent Systems Science and Engineering, Hubei Minzu University, Enshi 445000,
People's Republic of China }

\author{Jun Chen}
\affiliation{ College of Intelligent Systems Science and Engineering, Hubei Minzu University, Enshi 445000,
People's Republic of China }

\author{De-fu Hou}
\email{houdf@mail.ccnu.edu.cn} \affiliation{Key Laboratory of Quark and Lepton Physics
(MOE) and Institute
of Particle Physics, Central China Normal
University,Wuhan 430079,People's Republic of China}

\date{\today}
\begin{abstract}
In this study,  we derive the longitudinal and transverse gluon self-energies and the corresponding dielectric functions for a nonextensive QGP, based on nonextensive statistical mechanics and a kinetic theory framework. The nonextensive parameter $q$ enters these quantities primarily through the modification of the Debye mass. Utilizing the derived dielectric functions, we then calculate the collisional energy loss for a heavy quark using two established formalisms: the plasma physics-based Thoma-Gyulassy formula and the thermal field theory-originated Kirzhnits-Thoma formula.
Our results show that for both formalisms, the collisional energy loss increases with the nonextensive parameter
$q$ with this enhancement being more significant at higher incident quark momenta and suppressed for a heavier quark mass. The energy loss predicted from the Kirzhnits-Thoma formula is substantially larger than that from the Thoma-Gyulassy formula, and the nonextensive effect on the energy loss is more pronounced in the former. Furthermore, the mass suppression of the nonextensive effect on the energy loss is weaker in the Kirzhnits-Thoma approach. These calculations demonstrate that nonextensive statistics can significantly alter the  energy loss in the QGP.
\end{abstract}
\pacs{12.38.Mh}

\maketitle
\section{Introduction}
Quark-gluon plasma (QGP) is a special state of matter which is believed to be formed in ultrarelativistic heavy ion collisions in terrestrial laboratories. High energy heavy quarks produced by  hard scatterings in initial stage in heavy ion collisions traverse the created QCD plasma and degrade due to energy loss by collisions with medium partons or by bremsstrahlung of gluons, which is the so-called jet quenching phenomenon.
The energy loss can be directly related to experimental observables of ultrarelativistic heavy-ion collisions, such as nuclear modification factor, elliptic flow and azimuthal correlations, etc. The jet quenching has been regarded as one of the potential signals for the QGP formation in ultrarelativistic heavy ion collisions\cite{gyulassy90} and has attracted  persistent interest  in recent years\cite{gyulassy90,bjorken82,thoma91npb,mrow1991,koike1991,thoma91a,thoma91b,loss1,mustafa05,mazumder05,
romatschke04,romatschke05,peng24,peng25,du24,manuel21,comadran23,lou25,loss2,baier,qin15,
djord06,khzrzeev01,baier97a,baier97b,qin08,wicks07,wicks25,guo24,djor06,ayala08,cai25,wickphd,gossiauxprc78,
deja4,carrington17,jiang15,jiang17,jamal19,debnath,adil07,ghosh23,jamal21b,marjan24,thoma00a,thoma00b}.

Bjorken first investigated the energy loss suffered by a massless quark due to binary elastic scattering  off  thermal quarks and gluons in a thermal bath. The results are infrared and ultraviolet divergences in logarithmic form which are regulated by introducing the reasonable minimum and maximum momentum transfers $\mathfrak{q}_{min}$, $\mathfrak{q}_{max}$\cite{bjorken82}. Within the framework of  plasma physics, Thoma and Gyulassy have developed a formula for the collisional energy loss, in which the infrared divergence is automatically cut off due to the plasma collective effect\cite{thoma91npb}.
Some succedent investigations showed that the energy loss should be divided into two parts: soft-momentum transfer due to plasma collective effect and hard-momentum transfer due to individual collisions between the fast test parton and the medium constituents\cite{mrow1991,koike1991}.
A systematic thermal field theory framework including both the soft-momentum and hard-momentum transfers was proposed by Braaten and Thoma \cite{thoma91a,thoma91b}. In that method,  an intermediate momentum scale $gT\ll \mathfrak{q}^\ast \ll T$ is introduced. The hard thermal loop (HTL) resummation approach should be applied for the soft-momentum transfer $\mathfrak{q}<\mathfrak{q}^\ast$, while naive perturbative theory is used for the hard-momentum transfer $\mathfrak{q}>\mathfrak{q}^\ast$. The full energy loss is the sum of the two parts and the intermediate momentum scale $\mathfrak{q}^\ast$ is cancelled out\cite{thoma91a,thoma91b}. This method has been widely used for exploring the collisional energy loss in recent years\cite{mustafa05,mazumder05,wicks07,romatschke04,romatschke05,peng24,du24,manuel21,comadran23,lou25,peng25}.

Subsequent investigations showed that the radiative energy loss
dominates the collisional one and might be the main mechanism for the jet quenching\cite{loss1,loss2,baier,qin15}. However, Djordjevic et al found that it is not sufficient to explain the non-photonic electron spectrum  by  merely considering the radiative energy loss\cite{djord06}. Furthermore, the dead-cone effects\cite{khzrzeev01} and LPM effects\cite{baier97a,baier97b} severely reduce the radiative energy loss of the heavy quark. These facts  have attracted researchers to re-scrutinize the energy loss mechanism.
Some investigations reported that the collisional energy loss is comparable to or even larger than the radiative one in some energy regions\cite{mustafa05,mazumder05}. In addition, some authors argued that both the collisional energy loss and the radiative one are important and must be included simultaneously to interpret the experimental data\cite{qin08,wicks07,wicks25}. The study of the collisional energy loss has reawakened  people's interest in the heavy ion community\cite{wicks25,djor06,ayala08,deja4,carrington17,jiang15,jiang17,jamal19,debnath,adil07,ghosh23,jamal21b,
romatschke04,romatschke05,guo24,peng24,du24,manuel21,comadran23,lou25,peng25,cai25,wickphd,gossiauxprc78,marjan24,thoma00a,thoma00b}.

It should be noted that the parton systems produced in ultrarelativistic heavy ion collisions experience intrinsic fluctuations and display long-range interactions and correlations\cite{wilk2000,strongint4,mitra,osada,jiang25,qin25}.
The nonextensive statistics  which is the generalization of the Boltzmann-Gibbs statistics, becomes involved in dealing with such systems\cite{tsallis,tsallis09}. In the nonextensive statistics, entropy is not an additive quantity and the nonextensive parameter $q$ characterizes the entropy of a system\cite{tsallis,tsallis09,kapusta21}. When $q=1$, the nonextensive entropy recovers the usual Boltzmann-Gibbs entropy.  The deviation of $q$ from unit characterizes the extent of the nonextensivity of a system.
In recent years, there has been a great achievement in describing phenomenology of ultrarelativistic heavy-ion collisions
by applying the nonextensive statistics, such as
transverse momentum hadron
spectra\cite{wong,zheng0,wong13,liu14,zheng2,cleymans15,wong2,rath20j,zhen21,plb2024,plb2025,azmi,biro2009,cleymans12epja,cleymans12jpg,
cleymans18epja,cleymans13,shen19epja,epja53102,zheng1,biro20,zhen22,tang,cleymans16},
radial flow\cite{biro20,zhen22,tang,cleymans16}, etc.
Impressively, the nonextensive statistics has reproduced the charged hadron spectra the exponential form at low transverse momentum and the power-law tail at high transverse momentum perfectly.
It is amazing that the nonextensive statistics can describe the experimental data of spectra covering 14 or 15 orders of magnitude from the lowest to the considered highest transverse momentum by using only three input parameters\cite{wong,zheng0,kapusta21}.
Besides the experimental observables, there have been  considerable  issues in studying the nonextensive effect on the QGP properties by applying dynamical models associated with nonextensive distributions in recent years
\cite{osada,zheng24,biroepja2012,strickland22epjc,biro12prc,rath20e,sarwar,rath24,rath23e,singh25,cleymans16a,biro05prl,
deppman23,walton00prl,physa2023,megias2023,deppman24,baptista25,Bhattacharyya25,zhang24,zhang25}.

The study of the collisional energy loss of the heavy quark in the QGP  has mainly been conducted within the framework of standard Boltzmann-Gibbs statistics.
How the potential nonextensive nature of the QGP characterized by long-range interactions and memory effects, manifested itself in the energy loss mechanisms of propagating heavy quarks,
remains a significant open question.
Despite there are some Refs.\cite{walton00prl,physa2023,megias2023,deppman24,baptista25,cleymans16a,Bhattacharyya25,singh26} which have reported  the nonextensive effect on the nuclear modification factor $R_{AA}$, drag coefficient and diffusion coefficient within the framework of nonextensive Boltzmann equation/Fokker-Planck equation/Plastino-Plastino equation,
to the best of our knowledge, there are no prior direct studies about the effect of nonextensivity on the collisional energy loss  in the QGP available in the literature.
Though the collisional energy loss correlates with the drag coefficient, the precise determination of the energy loss and the related transport coefficients still lags behind\cite{xu99}.
Furthermore, it is  argued that Fokker-Planck and Langevin approaches may not capture the whole evolution properties of heavy quarks\cite{peng24,xu99}.
It is of great significance to study the heavy quark collisional energy loss  in terms of dynamical models combined with the nonextensive statistics.

In the present paper, we will investigate the nonextensive effect on the collisional energy loss suffered by a heavy quark traveling through the QGP within the frameworks of plasma physics method\cite{thoma91npb} and the finite temperature field theory\cite{thoma00a,thoma00b}. In these two methods, the collisional energy losses are determined by the longitudinal and transverse dielectric functions.
By applying the nonextensive distribution functions to the polarization tensor  derived from the QGP kinetic theory\cite{thoma2000,guoyun2023,strickland03}, we can obtain the longitudinal and transverse gluon self-energies, through which the longitudianl and transverse dielectric functions  can be derived. Through the sequent derivation, the nonextensive parameter $q$ is embedded in the gluon self-energies and the  dielectric functions. Therefore, one can study the nonextensive effect  on the collisional energy loss.

The paper is organized as follows. In Sec.\ref{frame}, we focus on the theoretical frameworks for the collisional energy loss
of the heavy quark crossing the QGP. In Sec.\ref{frame}A we introduce the Thoma-Gyulassy formula for the collisional energy loss derived from the plasma physics method.
In Sec.\ref{frame}B, we briefly review the Kirzhnits-Thoma formula for the energy loss which is derived from the Leontovich relations in thermal field theory.
In Sec.\ref{neem}, we briefly review several mainly used distribution functions in the nonextensive statistics. Then,
we derive the longitudinal and transverse gluon self-energies and the corresponding dielectric functions in terms of the polarization tensor obtained from the QGP kinetic theory associated with the nonextensive distribution functions.   The numerical results for the collisional energy loss of the heavy quark are presented in Sec.\ref{numerical}. Sec.\ref{summary} is conclusion and discussion.

The natural units $k_B=\hbar=c=1$, the metric $g_{\mu\nu}=(+,-,-,-)$ and  the following notations $P=(p^0,\textbf{p})$  with $p=|\textbf{p}|$ are used in the paper.

\section{The theoretical frameworks for the  collisional energy loss}\label{frame}

It should be noted that the collisional energy loss calculated with the systematic field theory method\cite{thoma91a,thoma91b} displays unphysical negative values at low incident momenta of the heavy quark\cite{romatschke04,romatschke05,peng24,guo24,djor06}.
In Ref.\cite{thoma91a}, the authors put forward an alternative approach  in which the resummation gluon propagators were used in the squared matrix element to evaluate the energy loss due to the soft-momentum transfer.
Djordjevic and Guo \et ~have developed that approach to arbitrary momentum transfer\cite{guo24,djor06}.
That intuitively consistent approach  enables people to address the full collisional energy loss including both the
soft-momentum and  hard-momentum contributions by extending the upper limit of momentum-integral in the expression of the collisional energy loss from
the intermediate momentum scale $\mathfrak{q}^\ast$ to the maximum momentum transfer $\mathfrak{q}_{max}$\cite{guo24,djor06}.
An advantage of that approach is that the unphysical negative values in low momentum region are eliminated\cite{guo24,djor06}.
Furthermore, it is no need to introduce the intermediate momentum scale $\mathfrak{q}^\ast$\cite{guo24,djor06} to distinguish between the soft and hard momenta. Some recent literature investigated the collisional energy loss in QCD plasma by following that approach\cite{wicks25,ayala08,cai25,wickphd,gossiauxprc78}.
In the present paper, we will investigate the full energy loss
within two frameworks which originally deal with the soft contribution of the energy loss  with the similar tactic.

\subsection{The plasma physics method: Thoma-Gyulassy formula}

When the fast heavy quark is introduced into the QGP, a chromoelectric field is induced in the medium. The color Lorentz force due to the induced chromoelectric field exerting in return on the travelling heavy quark  will cause the energy loss to the incident heavy quark. The formula of energy loss is given by\cite{thoma91npb}
\begin{equation}\label{enloss}
-\frac{dE}{dx}=-\frac{\textbf{v}}{v} q^a \cdot \rm Re \textbf{E}_{ind}^a(\textbf{x}=\textbf{v}t,t),
\end{equation}
where $\textbf{E}_{ind}^a(\textbf{x}=\textbf{v}t,t)$ is the induced chromoelectric field and $v=|\textbf{v}|$.   $q^a$ is the color charge relating to the fast quark and defined as  $q^aq^a=C_F\alpha_s$ with strong coupling constant $\alpha_s=g^2/4\pi$ and Casimir invariant $C_F=4/3$ for the fundamental representation. We assume that the QGP is a static medium and the running coupling constant is invariant.

Due to the external current of the fast quark $\textbf{j}^a_{ext}$,
the total chromoelectric field can be expressed in the momentum space in the linear response theory\cite{thoma91npb}
\begin{equation}\label{tote}
[\varepsilon_{ij}(\omega,k)-\frac{k^2}{\omega^2}(\delta_{ij}-\frac{k_ik_j}{k^2})]
E^a_{tot,j}(\omega,k)=\frac{4\pi}{i \omega}j^a_{ext,i}(\omega,k).
\end{equation}
Here, $\varepsilon_{ij}(\omega,k)$ is the dielectric tensor which reflects the chromoelectromagnetic properties of the QGP medium. In an isotropic and homogeneous medium, $\varepsilon_{ij}(\omega,k)$ can be decomposed into two components
\begin{equation}\label{diet}
\varepsilon_{ij}(\omega,k)=\varepsilon_L(\omega,k)\frac{k_ik_j}{k^2}+
\varepsilon_T(\omega,k) (\delta_{ij}-\frac{k_ik_j}{k^2}),
\end{equation}
\ie the longitudinal and transverse dielectric functions $\varepsilon_L(\omega,k)$, $\varepsilon_T(\omega,k)$.

The external current due to the fast quark  in the momentum space can be denoted as\cite{thoma91npb}
\begin{equation}\label{curr}
\textbf{j}^a_{ext}(\omega,k)=2\pi q^a \textbf{v} \delta(\omega-\textbf{k}\cdot \textbf{v}).
\end{equation}
According to Eqs.(\ref{tote})(\ref{diet})(\ref{curr}), by working out the induced chromoelectric field and substituting  it into Eq.(\ref{enloss}), we will arrive the  energy loss\cite{thoma91npb,mrow1991,koike1991}
\begin{eqnarray}\label{enlossm}
-\frac{dE}{dx}&=&-\frac{C_F\alpha_s}{2\pi^2v} \int d^3\textbf{k} \{\frac{\omega}{k^2}[ \rm Im \varepsilon_L^{-1}+
(v^2k^2-\omega^2)\\ \nonumber
 &\cdot& \rm Im (\omega^2\varepsilon_T-k^2)^{-1}]\}_{\omega=\textbf{k}\cdot \textbf{v}}.
\end{eqnarray}
That formula has been widely used to study the energy loss in the QCD plasma\cite{deja4,carrington17,jiang15,jiang17,jamal19,debnath,adil07,ghosh23,jamal21b,marjan24}.
It is clear that the longitudinal and transverse dielectric functions $\varepsilon_L(\omega,k)$, $\varepsilon_T(\omega,k)$ determine the energy loss.

\begin{figure}[t]
\begin{minipage}[h]{0.48\textwidth}
\centering{\includegraphics
{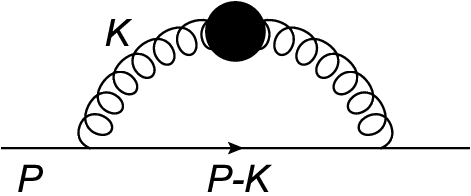}}
\end{minipage}
\caption{(color online) Feynman diagram for the heavy quark self-energy $\Sigma(P)$ with the resummaiton gluon propagator. } 
\end{figure}

\subsection{The Leontovich relations in thermal field theory: Kirzhnits-Thoma formula}

In quantum field theory, the collisional energy loss per unit length of a heavy quark can be expressed as\cite{thoma91a}
\begin{eqnarray}\label{1}
-\frac{dE}{dx}=-\frac{1}{v} \int (E-E')d \Gamma,
\end{eqnarray}
where $v$ is the velocity of the heavy quark, $E$ and $E'$ are energies of the heavy quark for the initial and final states.
$\Gamma$ is the interaction rate for the heavy quark which can be calculated with two methods. Firstly, $\Gamma$ can be derived from the matrix element for scattering of heavy quark $Q(P)$ off the medium parton $Q_i(L)$: $Q(P)+Q_i(L)\longrightarrow Q(P')+Q_i(L')$\cite{thoma91a}
\begin{eqnarray}\label{intme}
\Gamma&=&
\frac{1}{2E}\int \frac{d^3p'}{(2\pi)^32E'}\int \frac{d^3\textbf{l}}{(2\pi)^32 l}n(l)\int \frac{d^3\textbf{l}'}{(2\pi)^32l'}\\ \nonumber &\cdot&[1\mp n(l')]
(2\pi)^4\delta^4(P+L-P'-L') \frac{1}{2}\displaystyle\sum_{spins}|\mathcal{M}|^2.
\end{eqnarray}
Usually, interaction rate Eq.(\ref{intme}) has  been applied to Eq.(\ref{1}) to evaluate the collisional energy loss for the hard-momentum exchange\cite{mustafa05,mazumder05,wicks07,romatschke04,romatschke05,peng24,du24,manuel21,comadran23,lou25,peng25} in the systematic field theory method proposed by Braaten and Thoma \cite{thoma91a}.
Recently, some authors have applied Eq.(\ref{intme}) to (\ref{1}) with the hard thermal loop (HTL) resummation gluon or photon propagator in $\frac{1}{2}\displaystyle\sum_{spins}|\mathcal{M}|^2$ to
investigate the full energy loss including both soft and hard momentum transfers without the need to introduce the intermediate momentum scale to distinguish between the soft and hard momenta\cite{wicks25,guo24,djor06,ayala08,cai25,wickphd,gossiauxprc78}.

Secondly, the interaction rate can be evaluated by the heavy quark self-energy $\Sigma(P)$\cite{thoma91a}
\begin{eqnarray}\label{intme2}
\Gamma=
-\frac{1}{2E}[1-n_F(E)] \rm tr [(P\cdot\gamma+M) \rm Im\Sigma(E+i\epsilon,\textbf{p})].
\end{eqnarray}
In general,  Eq.(\ref{intme2}) has been applied to Eq.(\ref{1}) to deal with the collisional energy loss for the soft-momentum transfer in the Braaten-Thoma method\cite{thoma91a}. The Feynman diagram for the heavy quark self-energy  is shown in Fig.1 where the resummation gluon propagator, \ie the blob on the soft gluon line, should be used. The Feynman rules for the heavy quark self-energy in Fig.1 are
\begin{eqnarray}\label{qself}
\Sigma(P)=ig^2C_f\int\frac{d^4K}{(2\pi)^4}\Delta^{\mu\nu}(K)\gamma_\mu\frac{1}{(P-K)\cdot\gamma-M}\gamma_\nu.
\end{eqnarray}

Owing to Eqs.(\ref{1})(\ref{intme2})(\ref{qself}), by introducing the spectra functions $\rho_L,\rho_T$ according to the resummation gluon propagator $\Delta^{\mu\nu}(K)$, the energy loss can be expressed as\cite{thoma91a}
\begin{eqnarray}\label{enlossmm}
-\frac{dE}{dx}&=&\frac{C_F\alpha_s}{v^2} \int_{0}^{k^\ast} dkk \int_{-kv}^{kv} d\omega  \omega \{[ \rho_L(\omega,k)
\\ \nonumber &+&(v^2-\frac{\omega^2}{k^2})\rho_T(\omega,k)]\}.
\end{eqnarray}
In terms of the relations between spectra functions and dielectric functions, after some cumbersome algebra, one can obtain the collisional energy loss from (\ref{enlossmm})\cite{thoma91a,thoma00b}
\begin{eqnarray}\label{enlossmf}
-\frac{dE}{dx}&=&-\frac{C_F\alpha_s}{\pi v^2} \int_{0}^{k^\ast} \frac{dk}{k} \int_{-kv}^{kv} d\omega  \omega \{[ \rm Im \varepsilon_L^{-1}\\ \nonumber &+&
(v^2k^2-\omega^2)\cdot \rm Im (\omega^2\varepsilon_T-k^2)^{-1}]\}.
\end{eqnarray}
In Refs.\cite{thoma91a,thoma00b},  the authors have performed a comprehensive derivation of the collisional energy loss for a heavy fermion passing through QED plasma. The derivation processes for the heavy quark are similar to the QED case. For more details please refer to Refs.\cite{thoma91a,thoma00b}.  Here, we have replaced the electromagnetic coupling $\frac{e^2}{4\pi}$ in expression of QED energy loss by the color factor and strong coupling constant $\alpha_sC_F$ to obtain the corresponding QCD formula. Braaten and Thoma concluded that  quantum field theory result Eq.(\ref{enlossmf}) is consistent with Eq.(\ref{enlossm}) derived from the plasma physics method\cite{thoma91a,thoma00b}.

In the following, we will apply  a generalized Kramers-Kronig relation, \ie Leontovich relation to investigate  the collisional energy loss in terms of Eq.(\ref{enlossmf}).
Kirzhnits first applied the Leontovich relation\cite{leontovich} to study the stopping of a fast particle in a plasma\cite{kirzhnits88} and neutrino energy loss in matter\cite{kirzhnits90}.
Later, Thoma has developed  the Leontovich relation in thermal field theory to study the collisional energy loss of an energetic electron or muon in an electron-positron plasma and  an energetic parton in the quark-gluon plasma\cite{thoma00a,thoma00b}. In our consideration, we will extend that approach to study the collisional energy loss of the heavy quark in QGP incorporating the nonextensive effect.

Similar to Refs.\cite{thoma00a,thoma00b}, we introduce
the medium response function $R(k_0,k)$ which is defined according to the longitudinal and transverse dielectric functions
\begin{eqnarray}\label{response}
R(k_0,k)&=&R_L(k_0,k)+R_T(k_0,k), \\ \nonumber
R_L(k_0,k)&=&-\frac{1}{\varepsilon_L(k_0,k)}, \  R_T(k_0,k)=\frac{k^2-k_0^2}{k^2-k_0^2\varepsilon_T(k_0,k)}.
\end{eqnarray}
The energy loss Eq.(\ref{enlossmf}) can be expressed in terms of the response function $R(k_0,k)$ with the help of the property $\rm Im R(-\omega)=-\rm Im R(\omega)$\cite{thoma00a,thoma00b}
\begin{eqnarray}\label{enlossmfr}
-\frac{dE}{dx}=C_F\alpha_s \int_{0}^{\mathfrak{q}^\ast} d\mathfrak{q}\mathfrak{q}\frac{2}{\pi}\int_{0}^{\infty} d\omega  \omega \frac{\rm Im R(\omega,\sqrt{\mathfrak{q}^2+\omega^2})}{\mathfrak{q}^2+\omega^2}.
\end{eqnarray}
Here, we restricted ourselves to the ultrarelativistic case $v=1$ and replaced $k$ by the exchanged gluon momentum $\mathfrak{q}$ with   $\mathfrak{q}=\sqrt{k^2-\omega^2}$.

The response function $R$ fulfills the Kramers-Kronig relation which comes from the principle of causality\cite{kirzhnits88,thoma00a,thoma00b}
\begin{equation}\label{kk}
R(k_0,k)= \widetilde{R}+\frac{2}{\pi}\int_{0}^{\infty} d\omega  \omega \frac{\rm Im R(\omega,k)}{\omega^2-k_0^2-i\epsilon},
\end{equation}
with $\widetilde{R}=\lim_{k_0\rightarrow \infty} \rm Re R(k_0,k)$.
In a medium with spatial dispersion or under a magnetic field that is relativistic in character, the relativistic causality condition should be imposed which will lead to a more restrictive dispersion relation than Eq.(\ref{kk})\cite{leontovich,dolgov82}. One can generalize the Kramers-Kronig relation by performing a Lorentz transformation from $\omega$ and $k$ to $\omega'$, $k'$ given in a system which moves with the velocity $\textbf{u}$ relative to the initial system\cite{leontovich,kirzhnits90,dolgov82,kirzhnits88,thoma00a,thoma00b}. By choosing the strongest restrictions $\textbf{u}\cdot \textbf{k}=0$ and $|\textbf{u}|=1$ and utilizing the fact that $R$ is dependent on the modulus of $\textbf{k}$ in a homogeneous and isotropic medium, one can obtain the Leontovich relation\cite{kirzhnits88,kirzhnits90,leontovich,dolgov82,thoma00a,thoma00b}
\begin{eqnarray}\label{leno}
R(k_0,\sqrt{k^2+k_0^2})&=&R_\infty\\ \nonumber&+&\frac{2}{\pi}\int_{0}^{\infty} d\omega  \omega \frac{\rm Im R(\omega,\sqrt{k^2+\omega^2})}{\omega^2-k_0^2-i\epsilon}.
\end{eqnarray}
Here $R_\infty$ is given by $R_\infty=\lim_{k_0\rightarrow \infty} \rm Re R(k_0,\sqrt{k^2+k_0^2})$.

It should be noted that the second term  on the right hand side of the Leontovich relation Eq.(\ref{leno})
is in accord with the $\omega$-integral in the energy loss  Eq.(\ref{enlossmfr})\cite{kirzhnits88,thoma00a,thoma00b}
\begin{eqnarray}\label{enlossmfrf}
I=\frac{2}{\pi} \int_{0}^{\infty} d\omega  \omega \frac{\rm Im R(\omega,\sqrt{\mathfrak{q}^2+\omega^2})}{\mathfrak{q}^2+\omega^2},
\end{eqnarray}
when we replace $k_0$ by $i\mathfrak{q}$ and $\sqrt{k^2+k_0^2}$ by $0$, \ie $k^2=\mathfrak{q}^2$ in Eq.(\ref{leno}).
Therefore, from the Leontovich relation, we can obtain\cite{kirzhnits88,thoma00a,thoma00b}
\begin{equation}\label{iresult}
I=R(i\mathfrak{q},0)-R_\infty.
\end{equation}
From Eqs.(\ref{iresult})(\ref{enlossmfrf})(\ref{response})(\ref{enlossmfr}),
we can evaluate the  collisional energy loss in terms of the  dielectric functions associated with their asymptotic behaviors in the limits of $k\rightarrow 0$ and $\omega\rightarrow \infty$.

\section {The gluon self-energies and the dielectric functions in a nonextensive QGP}\label{neem}

In this section, we will derive the hard loop gluon self-energies and the medium dielectric functions
by applying the nonextensive distribution functions to the gluon polarization tensor obtained  from the kinetic theory.

\subsection{The nonextensive distribution functions}

In the nonextensive statistics,  in the case of the vanishing chemical potential, the single particle distribution functions for fermions (quarks and antiquarks) and bosons (gluons)  in the QGP can be expressed as \cite{rath20e,sarwar,rahamanIJMPA,jiang25,zhang24,zhang25}
\begin{equation}\label{fermi}
n_{q,\bar{q}}(\textbf{l})=\frac{1}{\left[1+(q-1)\beta E_l\right]^{\frac{q}{q-1}}+1},
\end{equation}
\begin{equation}\label{bose}
n_g(\textbf{l})=\frac{1}{\left[1+(q-1)\beta E_l\right]^{\frac{q}{q-1}}-1}.
\end{equation}
$E_l=\sqrt{\mathbf{l}^2+m_{q,\bar{q},g}^2}$ is single particle energy with medium parton mass $m_{q,\bar{q},g}$ and momenta $\mathbf{l}$. $\beta$ is related to temperature $\beta=\frac{1}{T}$.
The nonextensive distribution functions are also considered as  nonequilibrium ones and the nonextensive parameter $q$ measures the extent of deviation from equilibrated thermal distributions\cite{strickland22epjc,osada,rath20e,sarwar,rath23e,rath24}. If $q=1$, the nonextensive distribution functions turn to the usual Fermi-Dirac and Bose-Einstein  distributions, and the entropy recovers the usual  additive one.

At high temperature, the  nonextensive  distributions Eqs.(\ref{fermi})(\ref{bose}) can approximate as\cite{biro2009,zheng1,cleymans16,biro05prl,cleymans18epja,shen19epja}
\begin{equation}\label{fer}
n_{q,\bar{q},g}(\textbf{l})={\left[1+(q-1)\beta E_l\right]^{-\frac{q}{q-1}}}.
\end{equation}
It is argued that these distribution functions can satisfy the consistency of thermodynamics\cite{biro2009,zheng1,cleymans16,cleymans12epja,cleymans12jpg,cleymans18epja,shen19epja}.
Some authors have discussed these two versions of nonextensive distribution in Refs.\cite{kapusta21,biro2009,cleymans16}.
In addition, some other nonextensive distribution functions
\begin{equation}\label{fer2}
n_{q,\bar{q},g}(\textbf{l})=\frac{1}{\left[1+(q-1)\beta E_l\right]^{\frac{1}{q-1}}\pm1}
\end{equation}
and its high temperature approximation ${\left[1+(q-1)\beta E_l\right]^{-\frac{1}{q-1}}}$
have been widely employed to study heavy ion phenomenology\cite{strickland22epjc,mitra,strongint4,sarwar,rath23e,rath24,rath20e}.
Kapusta argued that though the nonextensive single particle distribution functions have been not derived from QCD,
they can interpolate between an exponential at low transverse momentum, reflecting the thermal equilibrium, to a power law at  high transverse momentum, reflecting the asymptotic freedom of QCD\cite{kapusta21}.

\subsection{The gluon self-energies and the dielectric functions}

It should be stressed that a plasma  with nonextensive distributions (\ref{fermi})(\ref{bose}) signifies  a stationary, homogenous and colorless plasma state\cite{osada,biro12prc}.
The gluon polarization tensor reads\cite{thoma2000,jiang25,guoyun2023,strickland03}
\begin{equation}\label{poltensor}
\Pi^{ij}(K)=-g^2\int \frac{d^3\textbf{l}}{(2\pi)^3} v^i \frac{\partial f(\textbf{l})}{\partial l^m} [\delta^{mj}+\frac{k^mv^j}{\omega-\textbf{k}\cdot\textbf{\^v}+i0^+}],
\end{equation}
where  $f(\textbf{l})$ relates to the distribution functions of quark, antiquark and gluon $n_q(\textbf{l}), n_{\bar{q}}(\textbf{l}),n_g(\textbf{l})$ as
\begin{equation}\label{dis}
f(\textbf{l})=2N_c n_g(\textbf{l})+N_f(n_q(\textbf{l})+n_{\bar{q}}(\textbf{l})).
\end{equation}
The derivation of Eq.(\ref{poltensor}) in kinetic theory is applicative to arbitrary distribution functions $f(\mathbf{l})$\cite{strickland03,guoyun2023} so long as the space-time is homogeneous\cite{guoyun2023}.
The hard loop gluon self-energies calculated from Eq.(\ref{poltensor}) for both equilibrium and nonequilibrium QGP are exactly consistent with those calculated based on the QCD diagrammatic approach\cite{thoma2000}.

If the nonextensive parameter $q$ is independent of momentum, as the Fermi-Dirac and Bose-Einstein distribution functions, the  nonextensive distribution functions  Eqs.(\ref{fermi})(\ref{bose}) and Eq.(\ref{dis}) satisfy
\begin{equation}\label{diffdis}
\frac{\partial f(\textbf{l})}{\partial l^m}=\frac{l^m}{l}\frac{\partial f(\textbf{l})}{\partial l}=v^m\frac{\partial f(\textbf{l})}{\partial l}.
\end{equation}
According to Eq.(\ref{diffdis}), one can decouple  Eq.(\ref{poltensor}) into  the integral over $l$ and the integral over the solid angle $\Omega$
\begin{equation}\label{poltensordeco}
\Pi^{ij}(K)=m_{DE}^2\int \frac{d\Omega}{4\pi}v^iv^m \cdot [\delta^{mj}+\frac{k^mv^j}{\omega-\textbf{k}\cdot\textbf{\^v}+i0^+}].
\end{equation}
$m_{DE}^2$ is the square of Debye mass in the nonextensive QGP which is denoted as
\begin{equation}\label{debyemass}
m_{DE}^2=-\frac{g^2}{2\pi^2}\int_0^\infty l^2 dl \frac{\partial f(\textbf{l})}{\partial l}.
\end{equation}
When the usual Fermi-Dirac and Bose-Einstein distribution functions are used, one can obtain  the HTL gluon polarization tensor with the Debye mass
\begin{equation}\label{debyemassHTL}
m_D^2=\frac{(2N_c+N_f)}{6}g^2T^2,
\end{equation}
according to  Eqs.(\ref{debyemass}) and (\ref{poltensordeco}).

By substituting (\ref{fermi})(\ref{bose}) into  (\ref{poltensordeco})(\ref{debyemass}), after some algebra, we can obtain the longitudinal and  transverse gluon self-energies in light of tensor decomposition (\ref{diet})
\begin{eqnarray}\label{logselfenergy}
\Pi_L(K)
=m_{DE}^2 [-1+\frac{\omega}{2k}(\ln|\frac{\omega+k}{\omega-k}|-i\pi\theta(k^2-\omega^2))],
\end{eqnarray}
\begin{eqnarray}\label{transelfenergyfinal}
\Pi_T(K)
&=&\frac{m_{DE}^2}{2}[\frac{\omega^2}{k^2} +\frac{\omega(k^2-\omega^2)}{2k^3}\cdot(\ln|\frac{\omega+k}{\omega-k}|\\ \nonumber&-&i\pi\theta(k^2-\omega^2))].
\end{eqnarray}
According to Eqs.(\ref{logselfenergy})(\ref{transelfenergyfinal}),
we obtain the longitudinal and transverse dielectric functions
\begin{eqnarray}\label{logdie}
\varepsilon_L(K)&=&1-\frac{\Pi_L(K)}{k^2}\\ \nonumber
&=&1+\frac{m_{DE}^2}{k^2}[1-\frac{\omega}{2k}(\ln|\frac{\omega+k}{\omega-k}|-i\pi\theta(k^2-\omega^2))],
\end{eqnarray}
\begin{eqnarray}\label{trandie}
\varepsilon_T(K)&=&1-\frac{\Pi_T(K)}{\omega^2}
=1-\frac{m_{DE}^2}{2k^2}[1+\frac{(k^2-\omega^2)}{2\omega k} \\ \nonumber &\cdot&
(\ln|\frac{\omega+k}{\omega-k}| -i\pi\theta(k^2-\omega^2))].
\end{eqnarray}

We have briefly reviewed the determination  of the  gluon self-energies $\Pi_L(K)$, $\Pi_T(K)$ and the dielectric functions $\varepsilon_L(K)$, $\varepsilon_T(K)$ according to the polarization tensor derived from kinetic theory associated with the nonextensive distribution functions.
In terms of the derived nonextensive dielectric functions $\varepsilon_L(K)$, $\varepsilon_T(K)$
and Eq.~(\ref{enlossm}) and Eqs.(\ref{iresult})(\ref{enlossmfrf})(\ref{response}) and (\ref{enlossmfr}), we can study the nonextensive effect on the  heavy quark collisional energy loss.

\section{Numerical results}\label{numerical}

\begin{figure}[t]\label{debye}
\begin{minipage}[h]{0.48\textwidth}
\centering{\includegraphics[width=7.5cm,height=4.5cm]{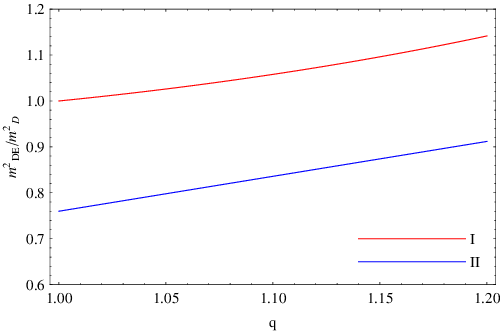}}
\end{minipage}
\caption{(color online) The scaled square of the Debye mass with respect to  the nonextensive parameter $q$. I(red curve): $\frac{m_{DE}^2}{m_D^2}$ calculated  with the nonextensive distribution functions (\ref{fermi})(\ref{bose}) without any approximation;  II(blue curve): $\frac{m_{DE}^2}{m_D^2}$ calculated with the  approximation (\ref{appnondis}).}
\end{figure}

In the following, we regard the nonextensive parameter $q$ as an input parameter to study its effect on the collisional energy loss of the heavy quark traversing the QGP. Some investigations showed that the consistency of the nonextensive
thermodynamics requires   $q>1$\cite{osada,cleymansprd} but $q<\frac{4}{3}$ \cite{cleymansprd,epja53102}.
Cleymans \et,  found  that to fit the RHIC and the LHC experimental data,  $q$ deviates from unit by $20\%$ at most\cite{cleymans12jpg}.
Therefore, we will study the heavy quark collisional energy loss in the QGP with some explicit values of $q$ in range $q\in[1,1.2]$. At the same time, we  assumed that  the masses of the charm and bottom quarks
$M_c=1.5\rm GeV$ and $M_b=5\rm GeV$ respectively.
In addition, $N_c=3$, $N_f=2$, $T=0.3\rm GeV$ and $\alpha_s=0.3$ are adopted to perform the numerical analysis.

Because  $\frac{\partial f(\textbf{l})}{\partial l^m}=v^m\frac{\partial f(\textbf{l})}{\partial l}$, as shown in Eq.(\ref{diffdis}),
the nonextensive parameter $q$ is embedded in  $m_{DE}^2$ in (\ref{debyemass}) through the distribution function $ f(\textbf{l})$ but has nothing to do with the integral over the solid angle $\Omega$ in Eq.(\ref{poltensordeco}).
Therefore, 
the nonextensive parameter $q$ modifies the Debye mass but does not has an impact on the structure of the polarization tensor.
The nonextensive longitudinal and transverse gluon self-energies Eqs.(\ref{logselfenergy})(\ref{transelfenergyfinal}) and the corresponding dielectric functions Eqs.(\ref{logdie})(\ref{trandie})  are the same as those in the HTL approximation  except that $m_D$ is replaced by $m_{DE}$.

By substituting the nonextensive distribution functions (\ref{fermi})(\ref{bose}) and (\ref{dis}) into (\ref{debyemass}), one can study the Debye mass numerically.
When $q\rightarrow 1$, the distribution functions (\ref{fermi})(\ref{bose}) approximate to the usual Fermi-Dirac and Bose-Einstein distribution functions,  the nonextensive  Debye mass recovers the result in the HTL approximation Eq.(\ref{debyemassHTL}).
We present $m_{DE}^2$ scaled by  $m_D^2$ as a function of $q$ numerically in Fig.2.
As shown the red curve, $m_{DE}^2$ achieves a monotonous increment as  the nonextensive parameter $q$ increases.

In our previous work\cite{jiang25}, we applied the nonextensive distribution functions expanded according to the Boltzmann distribution around $q=1$
\begin{eqnarray}\label{appnondis}
n_{q,\bar{q},g}=\textbf{e}^{-\beta E_l}+\frac{1}{2}(q-1)(-2+\beta E_l)\beta E_l\textbf{e}^{-\beta E_l}+...,
\end{eqnarray}
to the gluon polarization tensor (\ref{poltensordeco}) to derive the electric permittivity and the magnetic permeability to investigate the electromagnetic responses of the nonextensive QGP.
With that approximated distribution function (\ref{appnondis}) which also fulfills $\frac{\partial f(\textbf{l})}{\partial l^m}=v^m\frac{\partial f(\textbf{l})}{\partial l}$, we obtained  an analytical expression for the square of the Debye mass
\begin{equation}\label{debyemasse}
m_{DE}^2=\frac{2(1+(q-1))(N_c+N_f)}{\pi^2}g^2T^2,
\end{equation}
which is Eq.(26) in our previous work\cite{jiang25}. In addition, by using some different nonextensive distribution function (\ref{fer2}), Mitra obtained a complicated polynomial for the square of the Debye mass, for detailed arguments please refer to Eq.(72) in Ref.\cite{mitra}.

However, when $q=1$, Eq.(\ref{appnondis}) turns to the Boltzmann distribution function,  the approximation (\ref{appnondis}) omits some quantum effects. In the meantime, the analytical expression of the Debye mass (\ref{debyemasse}) reduces to $m_{DE}^2=\frac{2(N_c+N_f)}{\pi^2}g^2T^2$ which is much different from the HTL result $m_D^2$.
We also display results of the Debye mass (\ref{debyemasse}) with the blue curve in Fig.2 for comparison. As shown in Fig.2, the Debye mass calculated  with the nonextensive distribution functions (\ref{fermi})(\ref{bose}) without any approximation is much larger than that calculated in our previous work\cite{jiang25} with the  approximation (\ref{appnondis}).   It is argued that the collisional energy loss is sensitive to  the Debye mass\cite{koike1991}. As shown by the longitudinal and transverse dielectric functions (\ref{logdie})(\ref{trandie}), through $m_{DE}^2$
the nonextensive parameter $q$ influences the collisional energy loss in terms of Eq.(\ref{enlossm}) and Eqs.(\ref{enlossmfr})(\ref{enlossmfrf})(\ref{iresult})(\ref{response}).

It is argued that (\ref{enlossm}) and (\ref{enlossmf}) are applicative to the soft-momentum transfer\cite{mrow1991}.
Nevertheless, one can obtain the full energy loss within the $10\%$ level\cite{mrow1991,koike1991}
if the separation momentum scale $k^\ast$ is extended to the maximum momentum transfer determined by the kinematic constraint  of elastic scattering\cite{thoma91npb}
\begin{eqnarray}\label{kmax}
k_{max}=\frac{4Tp}{\sqrt{p^2+M^2}-p+4T},
\end{eqnarray}
where $p$,$M$ are the momentum and  mass of the incident heavy quark respectively.
A  recent report also supported that conclusion\cite{peng24}.
Recently, the maximum momentum transfer $k_{max}$ (\ref{kmax}) has been adopted in a number of references to employ the collisional energy loss\cite{peng24,adil07,wicks07,marjan24}.
In the present paper, we also apply that $k_{max}$ to the Thoma-Gyulassy formula to perform the numerical analysis of the energy loss.
\begin{figure}[t]
\begin{minipage}[h]{0.48\textwidth}
\centering{\includegraphics[width=7.5cm,height=4.635cm] {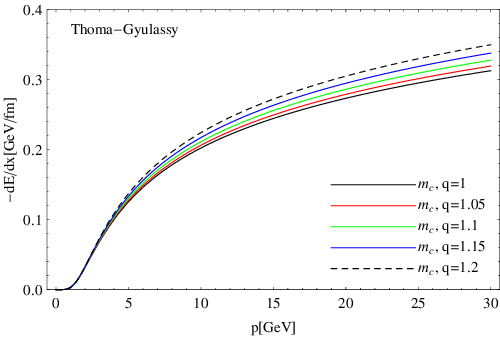}}
\end{minipage}
\begin{minipage}[h]{0.48\textwidth}
\centering{\includegraphics[width=7.5cm,height=4.5cm] {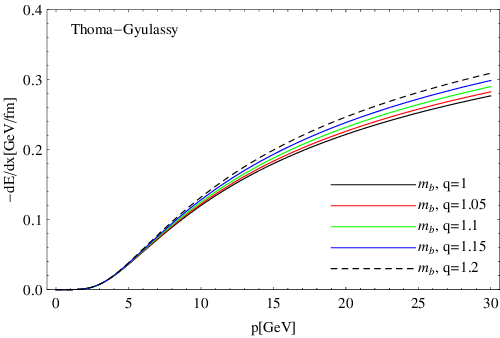}}
\end{minipage}
\caption{(color online) The collisional energy loss of a heavy quark in a nonextensive QGP in the Thoma-Gyulassy formula. The solid-black, red, green, blue and dashed-black curves are for the cases of $q=1, 1.05, 1.1, 1.15, 1.2$ respectively. Top panel: charm quark, Bottom panel: bottom quark. } \label{magnetic}
\end{figure}

\begin{figure}[t]
\begin{minipage}[h]{0.48\textwidth}
\centering{\includegraphics[width=7.5cm,height=4.635cm] {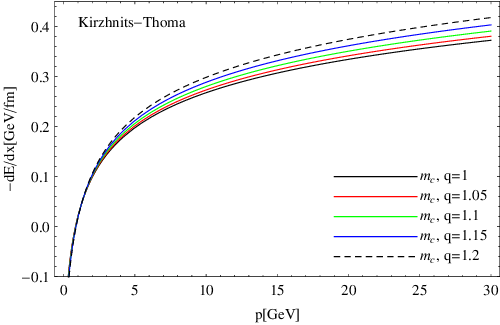}}
\end{minipage}
\begin{minipage}[h]{0.48\textwidth}
\centering{\includegraphics[width=7.5cm,height=4.5cm] {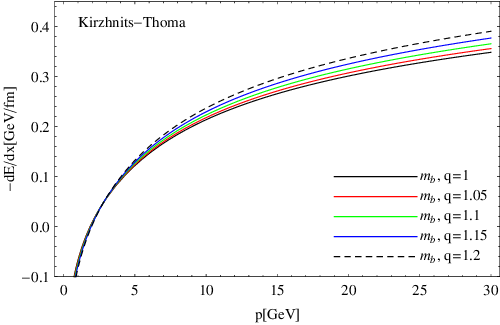}}
\end{minipage}
\caption{(color online) The collisional energy loss of a heavy quark in a nonextensive QGP in the Kirzhnits-Thoma formula. The solid-black, red, green, blue and dashed-black curves are for the cases of $q=1, 1.05, 1.1, 1.15, 1.2$ respectively. Top panel: charm quark, Bottom panel: bottom quark.}
\end{figure}

We plot the nonextensive collisional energy loss of the heavy quark calculated in the Thoma-Gyulassy formula in Fig.3. The solid-black, red, green, blue and dashed-black curves are for the cases of $q=1, 1.05, 1.1, 1.15, 1.2$, respectively. The top panel is for the charm quark, while the bottom panel is for the bottom quark. In the case of $q=1$, the dielectric functions $\varepsilon_L(K)$, $\varepsilon_T(K)$ recover the corresponding results in the HTL approximation, and the energy loss in that case has been addressed in Ref.\cite{thoma91npb}.
As shown in the top panel in Fig.3,  the collisional energy loss increases as $q$ changes  from $q=1$ to $q=1.2$. There is a critical momentum $p_c\sim 4\rm GeV$ below which all the nonextensive energy loss curves superpose each other and show no appreciable distinction among the different nonextensive curves, which indicates a trivial nonextensive effect on the collisional energy loss. When $p>p_c$,  the nonextensive effect on the collisional energy loss becomes pronounced as the incident momentum $p$ increases.
For the charm quark with $p=30\rm GeV$, the  energy loss for $q=1.2$ is roughly $13\%$ larger than that in the case of HTL approximation $q=1$.

As shown in the bottom panel in Fig.3, the collisional energy loss of the bottom quark
shows a similar $q-$dependence as that of the charm quark.  While the critical momentum for the bottom quark turns to a larger value $p_b\sim 8\rm GeV$.
For the bottom quark with $p=30\rm GeV$, the energy loss for $q=1.2$ is about $11.5\%$ larger than that in the case of  $q=1$. Furthermore, the nonextensive effect on the collisional energy loss of the bottom quark is less significant than that of the charm quark. These facts mean that the mass of the heavy quark suppresses the nonextensive effect  on the collisional energy loss.

The dielectric functions (\ref{logdie})(\ref{trandie}) satisfy $\varepsilon_L(k^0,0)=\varepsilon_T(k^0,0)$, which  leads to
\begin{eqnarray}\label{zerom}
R(i\mathfrak{q},0)=0
\end{eqnarray}
owning  to the fact there is no preferred directions in the medium at vanishing momentum\cite{thoma00a,thoma00b,kirzhnits88,dolgov82}. According to Eqs.(\ref{logselfenergy})(\ref{transelfenergyfinal}) (\ref{logdie}) and (\ref{trandie}), we can obtain  the dielectric functions in the high frequency limit\cite{thoma00a,thoma00b,kirzhnits88}
\begin{eqnarray}\label{trandielim}
\lim_{k_0\rightarrow \infty} \varepsilon_L(k_0,\sqrt{\mathfrak{q}^2+k_0^2})&=&1, \\ \nonumber
\lim_{k_0\rightarrow \infty} \varepsilon_T(k_0,\sqrt{\mathfrak{q}^2+k_0^2})&=&1-\frac{\omega_0^2}{k_0^2},
\end{eqnarray}
where
\begin{eqnarray}\label{trandielimww}
\omega_0^2=\lim_{k_0\rightarrow \infty} \Pi_T(k_0,\sqrt{\mathfrak{q}^2+k_0^2})=\frac{m^2_{DE}}{2}.
\end{eqnarray}
According to (\ref{response}) (\ref{iresult}) and the asymptotic behaviors of the dielectric functions in the high-frequency and zero-momentum limits (\ref{zerom})(\ref{trandielim}), we obtain the $I-$integral \cite{thoma00a,thoma00b,kirzhnits88}
\begin{eqnarray}\label{ifinal}
I=-R_\infty=\frac{\omega_0^2}{\mathfrak{q}^2+\omega_0^2}.
\end{eqnarray}

In terms of  (\ref{ifinal})(\ref{enlossmfrf}), the energy loss (\ref{enlossmfr}) turns to
\begin{eqnarray}\label{enleon1}
-\frac{dE}{dx}=C_F\alpha_s \int_{0}^{\mathfrak{q}^\ast} d\mathfrak{q}\frac{\mathfrak{q}\omega_0^2}{\mathfrak{q}^2+\omega_0^2}
=C_F\alpha_s\omega_0^2\ln\frac{\mathfrak{q}^\ast}{\omega_0},
\end{eqnarray}
where we have assumed that the intermediate momentum scale $\mathfrak{q}^\ast\gg\omega_0$. That result agrees up to a color factor and change in coupling constant with Eq.(62) in Ref.\cite{thoma91a} which is the soft contribution to the collisional energy loss in the limit $v\rightarrow 1$.

By extending $\mathfrak{q}^\ast$ to the maximum momentum transfer $\mathfrak{q}_{max}$ \cite{thoma00a,thoma00b}, one can obtain the Kirzhnits-Thoma formula for
the total collisional energy loss  from the  Leontovich relations
\begin{eqnarray}\label{enleon}
-\frac{dE}{dx}=C_F\alpha_s\omega_0^2\ln\frac{\mathfrak{q}_{max}}{\omega_0}.
\end{eqnarray}
If  $C_F\alpha_s$ is replaced by the electromagnetic coupling $\frac{e^2}{4\pi}$, (\ref{enleon}) will turn to the QED correspondence of the Kirzhnits-Thoma formula\cite{thoma00a,thoma00b}. In Refs.\cite{thoma00a,thoma00b}, based on  Eq.(\ref{enleon}), by adopting the reasonable maximum momentum transfer $\mathfrak{q}_{max}$ and $\omega_0$, the authors have compared the total energy loss results from the Kirzhnits-Thoma formula with those calculated from systematic field theory method\cite{thoma91a}.

\begin{figure}[t]
\begin{minipage}[h]{0.48\textwidth}
\centering{\includegraphics[width=7.5cm,height=4.635cm] {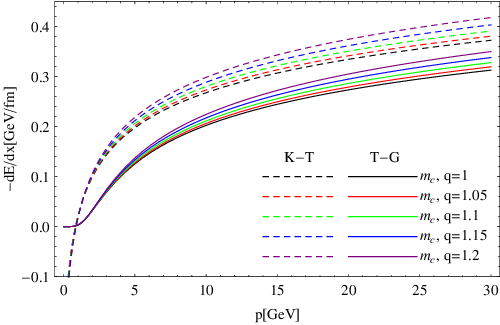}}
\end{minipage}
\begin{minipage}[h]{0.48\textwidth}
\centering{\includegraphics[width=7.5cm,height=4.5cm] {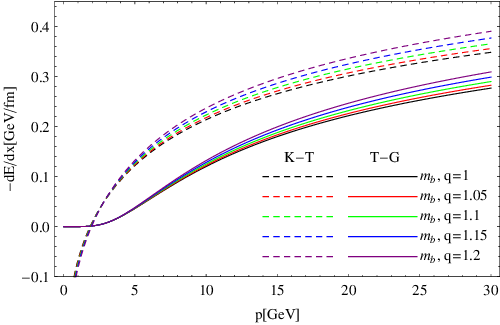}}
\end{minipage}
\caption{(color online) The collisional energy loss in a nonextensive QGP in the Thoma-Gyulassy formula (T-G) and in the  Kirzhnits-Thoma formula (K-T). The solid curves are for the case of T-G, while the dashed curves are for the case of K-T.
Top panel: charm quark, Bottom panel: bottom quark.} \label{magnetic}
\end{figure}

Fig.4 demonstrates the collisional energy loss of the heavy quark calculated with the Kirzhnits-Thoma formula by using the same maximum momentum transfer (\ref{kmax}). As shown in Fig.4,  the collisional energy losses have a similar dependence on the nonextensive parameter $q$, the incident momentum $p$ and the mass of the heavy quark as those calculated with the Thoma-Gyulassy formula.
However,  the critical momenta for both the charm and bottom quarks are around $3-4\rm GeV$, which implies that unlike the Thoma-Gyulassy case, the critical momentum is not sensitive to the mass of the heavy quark.

As discussed in  Sec.\ref{frame}B, for the soft colllisional energy loss, the quantum field theory formula Eq.(11) is equivalent to the Thoma-Gyulassy formula Eq.(\ref{enlossm}) derived from the plasma physics method\cite{thoma91a,thoma00b}.
To the best of our knowledge, there are no further studies about the comparison of the energy loss between the Thoma-Gyulassy formula and the Kirzhnits-Thoma formula when the maximum momentum transfer $k_{max}$  is adopted in those two formulas.
The comparison might provide some deep insights into the nonextensive effect on the energy loss. We present the numerical results from the  Thoma-Gyulassy  (abbreviated as T-G in the figure) and  the Kirzhnits-Thoma (abbreviated as K-T in the figure) formulas in the same panels in Fig.5.
The solid curves are for T-G and the dashed curves are for K-T, respectively.
Both for the charm and bottom quarks, the energy losses in the Kirzhnits-Thoma formula are much larger than those in the Thoma-Gyulassy formula.The nonextensive effect on the energy loss in the Kirzhnits-Thoma formula is  more remarkable than that in the Thoma-Gyulassy formula.  The gap of the energy loss between the Kirzhnits-Thoma  and Thoma-Gyulassy formulas for the bottom quark is wider than that for the charm quark.

To derive Eq.(\ref{enlossmfr}), we have applied the ultrarelativistic condition  $v=1$.
While in the Thoma-Gyulassy formula, the heavy quark has a finite velocity determined by the  incident momentum  which is smaller than $1$.
In addition, we have applied the asymptotic behavior of  the dielectric functions $\varepsilon_L(K)$, $\varepsilon_T(K)$ in the limits of $k\rightarrow 0$ and $\omega\rightarrow\infty$ to derive the Kirzhnits-Thoma formula Eq.(\ref{enleon}).
However, in the Thoma-Gyulassy formula, \ie (\ref{enlossm}),  finite $\omega,k$ are used in the dielectric functions $\varepsilon_L(K)$, $\varepsilon_T(K)$ to perform the numerical analysis.
Those considerations in two formulas lead to differences of the energy loss between the Kirzhnits-Thoma and Thoma-Gyulassy formulas.

\begin{figure}[t]
\begin{minipage}[h]{0.48\textwidth}
\centering{\includegraphics[width=7.5cm,height=4.635cm] {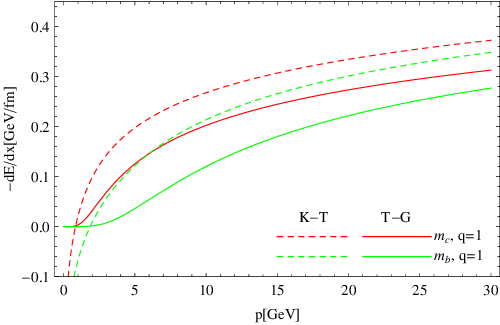}}
\end{minipage}
\caption{(color online) The collisional energy loss in a nonextensive QGP in the Thoma-Gyulassy formula (T-G) and in the  Kirzhnits-Thoma formula (K-T) for $q=1$. The solid curves are for the case of T-G, while the dashed curves are for the case of K-T. At the same time, the red curves are for the charm quark and the green curves are for the bottom quark.}
\end{figure}

There is a consensus that the charm quark loses more energy than the bottom quark. To elaborated the mass effect on the nonextensive energy loss, we present the collisional energy losses for both the charm and bottom quarks in the  Thoma-Gyulassy and Kirzhnits-Thoma formulas for $q=1$  in the same panels in Fig.6 for illustration.
The gap of the two dashed curves, reflecting the difference of energy loss between the charm and bottom quarks in the K-T case,  is narrower than that of the two solid curves, denoting the difference of energy loss between the charm and bottom quarks in the T-G case. That discrepancy is also appreciable in Fig.3 and Fig.4.
Therefore, compared to the Thoma-Gyulassy method,
the mass suppression effect on the extensive effect on the energy loss is weaker in the
Kirzhnits-Thoma formula. Those facts contribute to the wider gap of the energy loss curves  between the Kirzhnits-Thoma and the Thoma-Gyulassy cases for the bottom quark.

It is  clear to see that in the Kirzhnits-Thoma formula, there appears unphysical negative values for the collisional energy loss at low incident momenta, as shown in Fig.4, Fig.5 and Fig.6.  Furthermore, when $p\rightarrow 0$, the collisional energy loss shows a singularity. The reasons are that too small  an incident momentum $p$ will lead to the maximum momentum transfer $\mathfrak{q}_{max}=\frac{4Tp}{\sqrt{p^2+M^2}-p+4T}<\omega_0$ which breaks the assumption $\mathfrak{q}_{max}\gg\omega_0$ and $\mathfrak{q}_{max}\rightarrow 0$ for $p\rightarrow 0$. While in the Thoma-Gyulassy formula,  the maximum momentum transfer $k_{max}$ regulates the ultraviolet divergence. The infrared singularity is automatically screened by plasma collective effect. According to (\ref{logdie}) (\ref{trandie}), one can obtain\cite{thoma91npb}
\begin{eqnarray}\label{resumm}
Im \varepsilon_L^{-1}&=&k^2Im\frac{1}{k^2-\Pi_L}=k^2Im\Delta_L^\ast,\\ \nonumber
Im(\omega^2\varepsilon_T-k^2)^{-1}&=&Im \frac{1}{\omega^2-k^2-\Pi_T}=Im\Delta_T^\ast,
\end{eqnarray}
Where $\Delta_L^\ast,\Delta_T^\ast$ are the resummation longitudinal and transverse propagators of gluon in Coulomb gauge.
It means that the HTL resummation gluon propagators are adopted in the collisional energy loss (\ref{enlossm}) in the Thoma-Gyulassy formula implicitly. In the thermal field theory language, the infrared divergence is screened by the resummation propagators. For details, please refer to Sec.III of Ref.\cite{thoma91npb}.

In Ref.\cite{thoma91npb}, the authors have performed the numerical analysis of the energy loss for both  light and heavy quarks in terms of the Thoma-Gyulassy formula  Eq.(\ref{enlossm}).   However, that framework does not allow for recoil of the scattered  quark, which becomes involved especially for the light quark\cite{koike1991,thoma91a}. In addition, it is obscure to define the energy loss of a light particle at relativistic temperature because under such condition it will split into collinear particles, for detailed arguments please refer to Ref.\cite{thoma91a}.
We focused on the energy loss of  the heavy quark where kinematics prevents splitting into collinear particles\cite{thoma91a}. Intuitively, the energy loss of the heavy quark at  large momentum with $v\rightarrow1$ asymptotically corresponds to the case of a high energy light quark whose velocity $v\rightarrow1$. Therefore, the thermal field theory-based Kirzhnits-Thoma  formula will shed light on the energy loss for a high energy light quark.
It is  argued that the Kirzhnits-Thoma formula can provide a rough estimation for the upper bound of the collisional energy loss of the heavy quark with a very simple expression\cite{kirzhnits88}.


\section{Conclusion and Discussion}
\label{summary}

In this work, we have investigated the influence of nonextensive statistics, characterized by the nonextensive parameter $q$, on the collisional energy loss of heavy quarks traversing the QGP. Our analysis is grounded in the derived longitudinal and transverse gluon self-energies and the corresponding dielectric functions, which incorporate the nonextensive effect via a modified Debye mass. The energy loss was evaluated using two established formalisms: the Thoma-Gyulassy formula from plasma physics and the Kirzhnits-Thoma formula derived from thermal field theory via Leontovich relations.

Our numerical results reveal a consistent qualitative trend across both formalisms: the collisional energy loss increases with the nonextensive parameter $q$. This enhancement becomes more significant for higher incident momenta of the heavy quark. Physically, this suggests that deviations from standard Boltzmann-Gibbs statistics, as encapsulated in a $q>1$ scenario, lead to a medium with altered screening properties, thereby increasing the interaction strength and consequently the energy dissipation. Conversely, we find that the mass of the heavy quark acts as a suppressing factor on this nonextensive effect. A heavier quark experiences a less pronounced relative increase in energy loss due to $q$
-modifications, likely due to its reduced coupling to the collective, momentum-dependent screening modes of the plasma.

A key quantitative finding is the substantial difference in the magnitude of energy loss predicted from the two formulas. For both charm and bottom quarks, the Kirzhnits-Thoma formalism yields significantly larger energy losses than the Thoma-Gyulassy approach. Furthermore, the nonextensive effect is more pronounced, and the mass suppression is weaker, within the Kirzhnits-Thoma framework. This discrepancy highlights the sensitivity of energy loss calculations to the underlying theoretical formulation and the treatment of the medium's response. The wider gap for the bottom quark between the two formulas underscores the importance of the heavy quark mass as a scaling variable in these different theoretical contexts.

It is important to acknowledge the limitations of our current approach. Both the Thoma-Gyulassy and Kirzhnits-Thoma formulas are formally derived for soft momentum transfers. Our application, following the strategy in Refs.\cite{guo24,djor06}, extends their domain by integrating up to the kinematic maximum to estimate the total collisional loss. While this provides a complete picture without an arbitrary separation scale, a more rigorous treatment would involve a systematic field-theoretic approach, such as the one developed by Thoma and Braaten\cite{thoma91a,thoma91b}, or other intuitively consistent frameworks\cite{guo24,djor06,wicks25,ayala08,cai25,wickphd,gossiauxprc78}, to seamlessly incorporate both soft and hard contributions. Our use of the simplified integrated method is a first step to explore the nonextensive effect, but future work should employ these more foundational methods to solidify the conclusions.

The findings presented here open several avenues for future research. First, a comprehensive understanding of the jet quenching in the nonextensive QGP necessitates the study of radiative energy loss. The interplay between collisional and radiative mechanisms under nonextensive statistics could lead to non-trivial modifications of the total energy loss and the nuclear modification factor $R_{AA}$. Second, our results should be connected to experimental observables. Predictions for the $q$-dependence of heavy-flavor meson suppression and azimuthal anisotropy could provide a potential link to experimental data, offering a novel perspective on possible nonextensive features in heavy-ion collisions. Finally, exploring the origin of the large discrepancy between the two formalisms used here, perhaps through a unified derivation, remains an important theoretical challenge. These directions will be the focus of our subsequent investigations.

\vspace{3mm}

{\bf Acknowledgment}
We thank Wen-bin Chang and Ren-hong Fang for helpful discussions. This work is partly supported by the National Natural Science Foundation of China (NSFC) under Grant Nos.\ 12265013, 12435009 and 12275104 and the National Key Research and Development Program of China under Contract No. 2022YFA1604900.

\end{document}